\documentclass[12pt]{article}
\usepackage{amssymb,slashed,latexsym,ifpdf,amsmath}
\newcommand{\ft}[2]{{\textstyle\frac{#1}{#2}}}

\def\K{K\"{a}hler}
\newsavebox{\uuunit}
\sbox{\uuunit}
    {\setlength{\unitlength}{0.825em}
     \begin{picture}(0.6,0.7)
        \thinlines
        \put(0,0){\line(1,0){0.5}}
        \put(0.15,0){\line(0,1){0.7}}
        \put(0.35,0){\line(0,1){0.8}}
       \multiput(0.3,0.8)(-0.04,-0.02){12}{\rule{0.5pt}{0.5pt}}
     \end {picture}}

\newcommand{\bbox}{\lower.2ex\hbox{$\Box$}}
\newcommand{\Nn}{N}

\csname @addtoreset\endcsname{equation}{section}

\newif\ifpdf
\ifx\pdfoutput\undefined
   \pdffalse
   \usepackage{cite}
 \else
   \pdfoutput=1
   \pdftrue
  \usepackage[pdftex]{hyperref}
\fi

\newcommand{\rf}[1]{(\ref{#1})}

\def\aD3{{\overline {\rm D3}}}
\def\be{\begin{equation}}
\def\ee{\end{equation}}
\def\ba{\begin{array}}
\def\ea{\end{array}}
\def\bea{\begin{eqnarray}}
\def\eea{\end{eqnarray}}

\def\K{K{\"a}hler}

\begin{document}

\begin{titlepage}
\vspace{.5cm}
\begin{center}
\baselineskip=16pt

\rightline{Dedicated to I. V. Tyutin anniversary}

\hskip 1cm

\vskip 0.8cm

{\Large {\bf Matter-coupled de Sitter Supergravity }}

\

\

\

   {\large  \bf Renata Kallosh }\\

\vskip 1 cm

 Stanford Institute of Theoretical Physics and Department of Physics, \\
 \vskip 2mm
  Stanford University, Stanford, CA
94305 USA

\end{center}


\vskip 2cm
\begin{center}
{\bf Abstract}

\

\end{center}

 {\small De Sitter supergravity describes interaction of supergravity with general chiral and vector multiplets as well as one nilpotent chiral multiplet.  The extra universal positive term  in the potential due to the nilpotent multiplet, corresponding to the  anti-D3 brane in string theory, supports de Sitter vacua in these supergravity models.
In the flat space limit these supergravity models include the Volkov-Akulov model with a non-linearly realized supersymmetry. The rules for constructing  pure de Sitter supergravity action  are generalized here in presence of other matter multiplets. We present a strategy to derive the complete closed form general supergravity action with a given \K\, potential $K$,  superpotential $W$ and vector matrix $f_{AB}$ interacting with a nilpotent chiral multiplet. It  has  the potential $V=e^K(|F^2 |+ |DW|^2 - 3 |W|^2)$,  where $F$ is a necessarily non-vanishing  value of the  auxiliary field of the nilpotent multiplet. De Sitter vacua are present under simple condition that $|F^2|- 3|W|^2>0$. A complete explicit action in the unitary gauge is  presented.
}   
   \vspace{2mm} \vfill \hrule width 3.cm
{\footnotesize \noindent e-mail:  kallosh@stanford.edu
 }
\end{titlepage}
\addtocounter{page}{1}
\newpage
\section{  Historical note}

I was fortunate to start my career in physics under the great influence of Igor Tyutin's work in early 70's. Igor in collaboration with  E. Fradkin and I. Batalin at the Lebedev Physical Institute has made series of groundbreaking discoveries about the nature of quantum field theories with local gauge symmetries: Yang-Mills theories and gravity. This field received a tremendous boost in 1971 with the publication of the paper by 't Hooft suggesting that gauge theories with spontaneous symmetry can be renormalizable \cite{'tHooft:1971rn}. However, for a complete proof of renormalizability and unitarity of these theories it was necessary to prove equivalence between the results obtained in renormalizable gauges, where unitarity was hard to establish, and the unitary gauge, where renormalizability was not apparent. The equivalence theorem was proven in 1972 in our paper with Igor Tyutin, using path integral methods  \cite{Kallosh:1972ap}. An  independent  combinatorial proof of this result was given a month later  by 't Hooft and Veltman   \cite{'tHooft:1972ue}, with a reference to our work  \cite{Kallosh:1972ap}.

One of the most significant works by Tyutin was a discovery of what is now known as BRST symmetry. T in BRST reflects his preprint  which was published in 1975 as a Lebedev Institute preprint \# 75-39, in Russian; it was translated to English only 33 years later \cite{Tyutin:1975qk}. In 2009 Tyutin received the Dannie Heineman Prize for Mathematical Physics with  Becchi,  Rouet, and  Stora. By that time it became clear that the BRST symmetry is one of the most significant tools of theoretical physics.

Gauge invariance continues playing a profound role in modern physics, including quantum gravity, supergravity and string theory. In this paper I will discuss recent progress in de Sitter supergravity. This theory describes de Sitter vacua with a positive cosmological constant and spontaneous breaking of a  local supersymmetry. One of the interesting aspects of these theories is a possibility to make various gauge choices fixing local supersymmetry. We will find that there is a class of unitary gauges where the theory is particularly simple. 

\section{Introduction}

The complete action of the  supergravity multiplet interacting with a nilpotent goldstino multiplet   was recently constructed in 
\cite{Bergshoeff:2015tra,Hasegawa:2015bza}. The action was  named `Pure de Sitter Supergravity' in \cite{Bergshoeff:2015tra}  since it has  maximally symmetric classical de Sitter solutions,  even in the absence of fundamental scalars in the theory. The action has a non-linearly realized  local supersymmetry.  In the flat space limit the action becomes that of  the Volkov-Akulov  (VA) theory \cite {Volkov:1973ix} 
 with a global non-linearly realized supersymmetry.

The recent interest to nilpotent supergravity in application to cosmology  was  raised in  \cite{Antoniadis:2014oya} for VA-Starobinsky supergravity and developed in \cite{Dudas:2015eha} in the context of the chiral scalar curvature superfield ${\cal R}$ subject to a certain superfield constraint. Important implications of the nilpotent multiplet on the bosonic action and on cosmology were explained there. However, a complete and general locally supersymmetric component action including fermions in this approach was not yet constructed.  

A general approach to supergravity with a nilpotent multiplet was proposed in  \cite{Ferrara:2014kva}  using the superconformal approach in the form developed in  \cite{Kallosh:2000ve}.  It is described in detail in the textbook \cite{Freedman:2012zz}, where also earlier references to superconformal derivation of supergravity were given. Using this approach  it was  possible to construct a complete locally supersymmetric supergravity action interacting with a nilpotent multiplet \cite{Bergshoeff:2015tra,Hasegawa:2015bza}. In particular, the difference between the pure dS supergravity actions in \cite{Bergshoeff:2015tra} and in \cite{Hasegawa:2015bza} is due to a  different choice of the superconformal gauges for the local Weyl symmetry, local $\mathbb{R}$ symmetry and local special supersymmetry. One can also derive de Sitter supergravity starting with a complex  linear goldstino superfield \cite{Kuzenko:2015yxa}.

 The interest to a nilpotent goldstino in an effective ${\cal N}=1$ supergravity is enhanced by the better understanding of the relation to superstring  theory, specifically to the KKLT construction of  dS vacua \cite{Kachru:2003aw,Bergshoeff:2015jxa,Kallosh:2015nia}.  The role of an anti-D3 brane placed on the top of  the O3 (or O7) plane at the tip of the warped throat is now better understood \cite{Kallosh:2015nia}.  The corresponding string theory  constructions of the nilpotent  goldstino and an associated spontaneous breaking of supersymmetry by the anti-D3 brane action may lead to a UV completion of the effective 4-dimensional de Sitter supergravity interacting with a nilpotent multiplet.
  
An additional reason for using a nilpotent multiplet in supergravity is to describe inflation  \cite{Antoniadis:2014oya,Ferrara:2014kva}. The nilpotent multiplet is particularly useful when constructing cosmological models  with dark energy and susy breaking which are in agreement with the Planck/BICEP data since they have a flexible level of gravitational  waves and a controllable level of susy breaking parameter \cite{Kallosh:2014via,Dall'Agata:2014oka}.

The newly discovered dS supergravity action \cite{Bergshoeff:2015tra}  in the limit to a flat space  leads to a goldstino model  in the  form of a
nilpotent chiral multiplet as known from \cite{rocek,Komargodski:2009rz,Kuzenko:2010ef}, which is equivalent to the original globally supersymmetric VA goldstino model  \cite {Volkov:1973ix}. 
An interesting feature of the pure de Sitter supergravity \cite{Bergshoeff:2015tra,Hasegawa:2015bza} is that in the unitary gauge where the local supersymmetry is gauge-fixed by the choice of the vanishing fermionic goldstino, the action  is 
\begin{equation}
 e^{-1}   {\cal L}|_{{\rm goldstino}=0}
=\frac{1}{2\kappa ^2} \left[ R(e,\omega(e )) -\bar \psi _\mu \gamma ^{\mu \nu \rho } D^{(0)}_\nu \psi _\rho
 +{\cal L}_{\rm SG,torsion} \right] +\frac{3m^2}{\kappa^2} -f^2 +\frac{m}{2\kappa^2}\bar \psi _{\mu } \gamma ^{\mu \nu }\psi _{\nu } 
 \label{GaugeFixed}
\end{equation}
in notation of \cite{Bergshoeff:2015tra}, where $f$ and $m$ are constants. The action \rf{GaugeFixed}
for ${\bf \Lambda} = f^2- 3m^2/\kappa^2 >0$ is a  pure dS supergravity with a positive cosmological constant. In case that $f\neq 0$  and ${\bf \Lambda} = f^2- 3m^2/\kappa^2 \leq 0$ we have an AdS or Minkowski vacuum with spontaneously broken supersymmetry.
For $f=0$ the action \rf{GaugeFixed} has a restored linearly realized local supersymmetry and becomes a well known AdS supergravity action \cite{Townsend:1977qa}.

The non-trivial nilpotent multiplet is consistent  only when its auxiliary field is not vanishing.
The nilpotency condition for the field $S^2(x, \theta)=0$ for $S(x, \theta)=s(x)  +\sqrt 2\,  \theta \psi_s(x)  + \theta^2 F_s(x) $  includes three equations in terms of the component fields, a scalar,  sgoldstino  $s(x)$, a fermion,   goldstino $\psi_s(x)$ and an auxiliary field $F_s(x)$
\be
s \, F_s - 2 \psi_s^2=0\, , \qquad  s \,  \psi_s=0 \, , \qquad s^2(x)=0
\label{3eqNil}\ee
There are 3 distinct possibilities to resolve the nilpotency condition : the first one,  with a non-vanishing auxiliary field,  leaves us with a fermion goldstino without a fundamental scalar.
 \be  
  F_s \neq 0,   \qquad \Rightarrow \qquad \psi_s\neq 0, \quad  s\neq 0\, ,   \qquad  s= {\psi_s^2\over 2F_s} \qquad {\rm solves \, \, all \, \, three \, \, equations}
  \ee
  This solution  is used in the locally supersymmetric action including a nilpotent multiplet without a fundamental scalar.
The second solution still with a  non-vanishing auxiliary field, has a solution where only the auxiliary field does not vanish, but both the scalar and the fermion vanish. 
 \be   
 F_s \neq 0, \qquad \Rightarrow \qquad  \psi_s=  0, \quad  s= 0 \hskip 3.2 cm  {\rm solves \, \, all \, \, three \, \, equations}
 \ee
 This solution is the one where the local supersymmetry is gauge-fixed in the unitary gauge with $\psi_s=0$ and $s=0$.
 
 The third possibility, with the vanishing  auxiliary field,  is a trivial solution, goldstino as well as sgoldstino both have to vanish, to solve all three equations \rf{3eqNil}
  \be 
\noindent F_s =  0  \qquad \Rightarrow \qquad s=\psi_s=0  \hskip 4 cm  {\rm solves \, \, all \, \, three \, \, equations}
\ee

The purpose of this paper is to outline the strategy for the derivation of  a complete matter-coupled supergravity action with one of the multiplets constrained to be a nilpotent one. The price for a relatively easy procedure of finding dS vacua with spontaneously broken supersymmetry is the non-linear fermion terms in the action. We will find, however, that all these terms can be presented in a closed form. The upshot of the result obtained in this paper is that we will describe a complete de Sitter supergravity with a non-linearly realized local supersymmetry, which has the following potential
\be
V= e^{K} ( F_s K^{s\bar s} \bar F_{\bar s} + D_i W K^{i\bar k} \bar D_{\bar k} \overline W- 3 W \overline W)\equiv e^K( |F^2 |+ |DW|^2) - 3m_{3/2}^2 \ ,
\label{pot}\ee
where $F_s\neq 0$ is the non-vanishing auxiliary field of the nilpotent multiplet.
For the  vacuum to have a positive cosmological constant it is sufficient to require that at the minimum
\be
e^K |F^2 |- 3m_{3/2}^2 >0 \ .
\ee
In the past de Sitter vacua in models without a nilpotent multiplet were constructed by requiring that $ |DW|^2 - 3m_{3/2}^2>0$, see for example \cite{Saltman:2004jh} and more references for such models  in string theory inspired supergravity in \cite{Kallosh:2015nia}. The common feature of all such models is that to provide the positive value of $|DW|^2 - 3m_{3/2}^2$ in the vacuum requires a significant engineering and practically always some computer computations. Several chiral multiplets are required and a choice of the \K\, potential and a superpotential has to be made carefully. The advantage of our de Sitter models with a nilpotent multiplet is that all we need is that one constant parameter, the value of $|F^2 |$ at the vacuum is larger that another constant parameter, the value of $3m_{3/2}^2$ at the vacuum. 

Thus we proceed with the strategy  to derive  de Sitter supergravity: general chiral and vector multiplets and a nilpotent multiplet, coupled to supergravity multiplet.

\section{Superconformal theory with a nilpotent multiplet}
We start with the underlying superconformal action in \cite{Ferrara:2014kva}: the $SU(2,2|1)$-invariant Lagrangian of ${\cal N}=1$ supergravity coupled to
 chiral multiplets $X^I$ and to Yang--Mills vector multiplets $\lambda^A$ 
superconformally.   It consists of  4 parts,
each of which is conformally invariant separately.  The superconformal action with all supersymmetries linearly realized is given by the following expression:
 \begin{equation}
{\cal L}_{sc} =  [\Nn (X^I,\bar X^{\bar I})]_D + [\mathcal{W}(X^I)]_F  + \left[ f_{AB } (X) \bar \lambda_L ^A \lambda_L ^B \right] _F + \left[\Lambda  \, (X^1)^2 \right] _F\,.
\label{symbLsc}
\end{equation}
The first 3 terms  are standard  and in this form are described in \cite{Kallosh:2000ve}  and in detail in the textbook \cite{Freedman:2012zz}.  We will use here the  notations of  \cite{Freedman:2012zz}. 
The non-standard last term depends on the Lagrange multiplier chiral superfield $\Lambda$. As different from other standard chiral superfields $X^I$,  it is not present in the \K\, manifold of the embedding space $\Nn (X^I,\bar X^{\bar I})$. Therefore it can be eliminated on its algebraic equations of motion.
The three functions $\Nn (X^I,\bar X^{\bar I})$, $\mathcal{W}(X^I)$, $ f_{AB } (X)$ as well as all  chiral superfields  transform in a
homogeneous way under local Weyl and $\mathbb{R}$ transformations:
Here $X^I$ include the chiral compensating multiplet    $\{X^0, \Omega^0, F^0\}$, a chiral goldstino multiplet  $\{X^1, \Omega^1, F^1\}$, generic matter multiplets 
$\{X^i, \Omega^i, F^i\}$  $i=2,...,n$ and a Lagrange multiplier multiplet $\{ \Lambda, \Omega^\Lambda, F^{\Lambda} \} $.
Equations of motion for the Lagrange multiplier multiplet $\Lambda$ consist of 3 component equations, for each of its components. One of these equations for $\Lambda(x)$ is \cite{Bergshoeff:2015tra}
\begin{equation}
   2 X^1 F^1 -\bar \Omega^1  P_L\Omega^1  +\sqrt{2}\bar \psi_\mu \gamma^\mu X^1  P_L\Omega^1 +\ft12 \bar \psi _{\mu }P_R \gamma ^{\mu \nu }\psi _{\nu }(X^1) ^2 =0\,.
 \label{fieldeqnLambda}
\end{equation} 
It  is solved  by
\begin{equation}
X^1 (x)  =\frac{\overline{\Omega^1 (x)}P_L \Omega^1(x) }{2F^1 (x)}\equiv \frac{(\Omega^1)^2 }{2F^1}\,, \qquad (X^1(x) )^2=0 \ ,
 \label{s}
\end{equation}
which also solves the remaining equations for $\Omega^\Lambda(x)$ and $F^\Lambda(x)$.

 The next step in producing supergravity action starting from the superconformal action \rf{symbLsc} is to impose the constraint \rf{s}, to 
 eliminate the auxiliary fields $F^I$ and to fix local symmetries of the superconformal action by gauge-fixing  local Weyl and $\mathbb{R}$ symmetry and a local special supersymmetry. The procedure is well known in the absence of the  constraint \rf{s}. 
The non-vanishing value of the auxiliary fields is according to eq. (17.21)  in \cite{Freedman:2012zz}
\be
 \bar F^{\bar I}_G = N^{ \bar I I }( -  {\cal W}_I  + {1\over 2} N_{I\bar K \bar L}  
\bar \Omega^{\bar K} \bar \Omega^{\bar L} + {1\over 4} f_{AB I} \bar \lambda^A P_L \lambda^B) \ .
\label{compl}\ee
 This is due to a Gaussian dependence of the action (17.19) on $F$ of the kind
 \be
 {\cal L}_{G} (F)=  N_{I\bar J} F^I \bar F^{\bar J} + [F^{I}  (  {\cal W}_I  - {1\over 2} N_{I\bar K \bar L}  
\bar \Omega^{\bar K} \bar \Omega^{\bar L} - {1\over 4} f_{AB I} \bar \lambda^A P_L \lambda^B)   +h.c.] \ ,
 \ee 
or equivalently 
\be
 {\cal L}_{G} (F)=  N_{I\bar J} F^I \bar F^{\bar J} + F^{I}  \bar F_{I\, G}   +\bar F^{\bar I} F_{I\, G} \ .
 \ee

 The new situation when one of the chiral multiplets is nilpotent may be studied first using the superconformal framework in eq. \rf{symbLsc}. Note that 
 $N(X^I, \bar X^{\bar I})$,  $\mathcal{W}(X^I)$ and $f_{AB } (X)$ are algebraic functions of $X^I$. The presence of the last term in \rf{symbLsc},   $\left[\Lambda  \, (X^1)^2 \right] _F$ suggests that $N(X^I, \bar X^{\bar I})$,  $\mathcal{W}(X^I)$ and $f_{AB } (X)$ can only depend on $X^1\bar X^{\bar 1}$ on $X^1$ and on $\bar X^{\bar 1}$, since according to \rf{s} $(X^1(x) )^2=0$.  We design the dependence on $X^1, \bar X^{\bar 1}$ in $N(X, \bar X)$ such that it depends only on $X^1\bar X^{\bar 1}$. 
 
 First, we will notice that in the expression $F^1 \bar F_{1 G} $ a dependence in $\bar F_{1 G} $ on $X^1$ will come as proportional to $F^1 X^1= {(\Omega^1)^2\over 2}$ which makes such terms  $F^1$ independent. Or if there are terms in $\bar F_{1 G} $ of the form  $X^1\bar X^{\bar 1}$, they also  become $F^1$-independent. Thus terms in ${\cal L}_{G} (F)$ which depend on $F^1 X^1 $ or $F^1 X^1 \bar X^{\bar 1}$ can be moved out from the $F$-dependent part of the action into an $F$-independent part.
 
 However, term in $\bar F_{1 G} $ depending on $\bar X^{\bar 1}$ cannot be removed that way. Thus we have to keep in mind that in $F^1 \bar F_{1G}$ we keep only terms linear in $\bar X^1$
 \be
 \bar F_{1 G}(\bar X^1) =  \bar F_{1 G}(\bar X^{\bar 1}=0)+ \bar F_{1 G, \bar1} \bar X^{\bar 1}
\label{barX} \ee
 and a conjugate. Using this fact one can look at the complete superconformal action in eq. (17.22) in \cite{Freedman:2012zz} and check that it can be given in the form
 \begin{equation}
e^{-1}{\cal L}= (F^I-F^I_{G})N_{I\bar I} (\bar F^{\bar I} - \bar F^{\bar I}_G) - F^I_{G}  N_{I\bar I}  \bar F^{\bar I}_G  + \bar X^1\, A_c\, X^1 + X^1\bar B_c + B_c\bar X^{\bar 1} +C_c\ .
\label{actionSC}
\end{equation}
We define
\be
 e^{-1}{\cal L}_X\equiv  - F^I_{G}  N_{I\bar I}  \bar F^{\bar I}_G  + \bar X^1\, A_c\, X^1 + X^1\bar B_c + B_c\bar X^{\bar 1} +C_c \ .
\ee
The part depending on auxiliary fields $ (F^I-F^I_{G})N_{I\bar I} (\bar F^{\bar I} - \bar F^{\bar I}_G)$ we present as  
\begin{equation}
 (F^1-F^1_{G})N_{1\bar 1} (\bar F^{\bar 1} - \bar F^{\bar 1}_G) + [(F^1-F^1_{G})N_{1\bar i} (\bar F^{\bar i} - \bar F^{\bar i}_G) +h.c. ] +  (F^i-F^i_{G})N_{i\bar k} (\bar F^{\bar k} - \bar F^{\bar k}_G) \ .
\label{actionF}
\end{equation}
Dependence on $F^i$ and $\bar F^{\bar i}$ is Gaussian since $ e^{-1}{\cal L}_X$ does not depend on these fields. We can integrate over them and find that \be
\bar F_i -  \bar F_{i G} =0
\label{i}\ee
 and get the  action in the form
\begin{equation}
e^{-1}{\cal L}= (F^1-F^1_{G})( N^{1\bar 1})^{-1} (\bar F^{\bar 1} -\bar F^{\bar 1}_G)  +e^{-1}{\cal L}_X\ .
\label{actionSC1}
\end{equation}
where 
\be
( N^{1\bar 1})^{-1}= N_{1\bar 1} - N_{1\bar k} N_{k\bar k} ^{-1} N_{k\bar 1} \ .
\label{inverse}\ee
The second term in eq. \rf{inverse} is proportional to $X\bar X$. Therefore, all terms in \rf{actionSC1} of the form  
$ (F^1-F^1_{G})(- N_{1\bar k} N_{k\bar k} ^{-1} N_{k\bar 1})(\bar F^{\bar 1} -\bar F^{\bar 1}_G)  $ add to ${\cal L}_X$ since $F^1 X^1= {\Omega^2/2}$. This appears to modify   ${\cal L}_X$ into $\tilde {\cal L}_X$. However, we may take into account that $F^1-F^1_{G}$ is non-vanishing only due to a non-Gaussian nature of the dependence of the action on $F^1$ and this difference is proportional to goldstino. But the term 
$ N_{1\bar k} N_{k\bar k} ^{-1} N_{k\bar 1}$ already has all possible dependnece on goldstino via $X\bar X$. Therefore these terms do not change the action.

Thus, the remaining expression depending on auxiliary field is
\begin{equation}
e^{-1}{\cal L}= (F^1-F^1_{G} ) N_{1\bar 1} (\bar F^{\bar 1} -\bar F^{\bar 1}_G)  +e^{-1} {\cal L}_X\ .
\label{actionSC2}
\end{equation}
One more step is required to understand the possible role of the terms in \rf{barX}. We rewrite the first term in \rf{actionSC2} as follows
\be
 (F^1-F^1_{G} )  (\bar F_{ 1} -\bar F_{ 1 G}) =  (F^1-F^1_{G} )  (\bar F_{ 1} -(N_{1\bar 1} \bar F^{\bar 1} _{ G}+ N_{1\bar k} \bar F^{\bar k} _{ G}) \ ,
\label{actionSC3}
\end{equation}
using \rf{i} and taking into account that
 $\bar F_{1\, G} = N_{1\bar 1} \bar F^{\bar 1} _{ G}+ N_{1\bar k} \bar F^{\bar k} _{ G} $ and according to discussion  around eq. \rf{barX} it is $X$-independent but may have some $\bar X$-dependence. Here we take into account that $N_{1\bar 1}$ is $X$ and $\bar X$ independent and that 
$N_{1\bar k} $ is proportional to $\bar X$. This means that linear dependence of $\bar F_{1\, G}$ on $\bar X$ translates into a linear dependence on $\bar X$ in $\bar F^{\bar 1} _{ G}$.

Now something interesting takes place: let us define $\bar F^{\bar 1}+ \bar F^{\bar 1} _{ G, \bar1} \bar X^{\bar 1}\equiv (\bar F^{\bar 1})'$.  It appears that the relevant dependence on $\bar X^{\bar 1}= {\bar \Omega^2\over 2 \bar F^{\bar 1}}= {\bar \Omega^2\over 2 (\bar F^{\bar 1})'}$ does not see the difference between $F$ and $F'$ since $\bar X^{\bar 1} \bar \Omega^2=0$. An analogous shift can be made with regard to $F^1$ which will absorb the term in $F^1_{G}$ depending on $X^1$.
This brings us to the following action
\begin{equation}
e^{-1}{\cal L}= (F^1-F^1_{G} ) N_{1\bar 1} (\bar F^{\bar 1} -\bar F^{\bar 1}_G)|_{X^1=0}  +e^{-1} {\cal L}_X\ .
\label{actionSC3a}
\end{equation}
One more step is useful to reduce the problem to the one which was already solved in \cite{Bergshoeff:2015tra}. $N_{1\bar 1}$ can depend on moduli and we would like to redefine $(F^1-F^1_{G} )$ by absorbing the factor $\sqrt {N_{1\bar 1}}$.
This brings us finally to the action in the form
 \begin{equation}
e^{-1}{\cal L}_{gen}= (F-F_{G0}) (\bar F - \bar F_{G0}) - F_{G0}  \bar F_{G0}  + \bar X\, A\, X + X\bar B + B\bar X +C\ .
\label{actionSC4}
\end{equation}
with $X= {\Omega^2\over 2 F}$. Here 
\be
F_{G0}= F_{G}|_{X^1=\bar X^1=0}
\ee
The explicit identification of all entries in eq. \rf{actionSC4} requires a significant effort. This will be done in a separate work \cite{RT}. Here we have pointed out the important steps and explained why we can make  the same construction as in the pure de Sitter case. The final action where the auxiliary field $F$ is eliminated in the symbolic form using all entries in  \rf{actionSC4} is very simple. It is given by
 \begin{equation}
e^{-1}{\cal L}_{gen}|_{{\delta {\cal L}\over \delta F}=0} =  - F_{G0}  \bar F_{G0} +C + \Big [   \bar X\, A\, X + X\bar B + B\bar X    -  {1\over F_{G0}\bar F_{G0}} |\bar X (A X+B) |^2 \Big ]_{X= {\Omega^2\over 2 F_{G0}} } \ ,
\label{actionSC5}
\end{equation}
or, equivalently, 
 \begin{equation}
 - F_{G0}  \bar F_{G0}  +   {\bar \Omega^2\over 2 \bar F_{G0} }\, A\, {\Omega^2\over 2 F_{G0} } - {\Omega^2\over 2 F_{G0} } \bar B - B {\bar \Omega^2\over 2 \bar F_{G0} }  +C   -  {1\over F_{G0} \bar F_{G0} } \Big |{\bar \Omega^2\over 2 \bar F_{G0} } \Big (A {\Omega^2\over 2 F_{G0} }+B\Big ) \Big |^2  \ .
\label{actionSC6}
\end{equation}
This action consist of the original action with $X^1$ replaced by ${\bar \Omega^2\over 2 \bar F_{G0} }$,  and 
higher order in spinors term given by the last term in eq. \rf{actionSC6}.

 The relation between supergravity moduli $z^{\alpha}, \alpha =1,...,n$ and the  superconformal variables $X^I$ is explained in the superconformal gauge in $\kappa^2=1$ units
  \be
N(X, \bar X) =-{3}\, ,  \qquad y=\bar y \, ,  \qquad  N_I \Omega^I=0  \ ,
\label{Eframe} \ee
 introduced in \cite{Kallosh:2000ve} is explained in detail  in  \cite{Freedman:2012zz}. One can start with the superconformal action derived here, gauge-fix it and derive the corresponding supergravity model where one of the multiplets in a nilpotent one.
 
  Alternatively, one can start with the known supergravity action in case of chiral and vector multiplets, and find the relevant modifications in case of one of the multiplets being a nilpotent one. This strategy will be described in the next section.
 
 \section{Supergravity action}
 
 A general supergravity action with chiral and vector multiplets derived from the superconformal one is given in sec. 18.1 in \cite{Freedman:2012zz}.
But the result we would like to present here is the deformation of the standard action in sec. 18.1 in \cite{Freedman:2012zz} for a given $K(z, \bar z)$, $W(z)$ and $f_{AB \alpha}$ due to the fact that the superconformal chiral superfield $X^1$ satisfies the constraint \rf{fieldeqnLambda}. We will assume that the moduli $z^{\alpha}$ with $\alpha=1,...,n$ are split into $z^1 = {X^1\over X^0}=S$ and the rest $z^i = {X^I\over X^0}$, $i=2,...,n$. We will study the class of models such that
 \be
 K(z^i, \bar z^{\bar i}; S, \bar S)= K(z^i, \bar z^{\bar i}) + g_{S\bar S}(z, \bar z) S\bar S \ ,
\label{K} \ee
  \be
 W(z, S)= g(z) +   S f(z) \, ,\qquad f_{AB}(z, S) = p_{AB}(z) + S \, q_{AB} (z)  \ .
\label{W} \ee
In view of the fact that $S^2=0$ both holomorphic functions $W$ and $f_{AB}$ depend only on linear functions of $S$. These are the most general form of such holomorphic functions,   as was recognized in \cite{Kallosh:2014via}, since  all powers of $S^n$  for $n>1$ vanish.  
The \K\, potential in most general case could have some linear dependence on $S$ and $\bar S$. We will not consider such models, our \K\, potential will depend on $S\bar S$.

The conceptual simplicity in the superconformal theory of the single nilpotent chiral multiplet is due to a   clear off-shell local supersymmetry transformation of the supermultiplets in the action \rf{symbLsc}. For example, for the off-shell chiral goldstino multiplet   $\{X^1, \Omega^1, F^1\}$ of the Weyl weight $\omega =1$ with the off-shell Lagrange multiplier $\{ \Lambda, \Omega^\Lambda, F^{\Lambda} \} $ the local Q-and S-supersymmetry rules is
\be
\delta \Omega^1= {1\over \sqrt 2} P_L (\slashed{D} + F^1)\epsilon + \sqrt 2 X^1 P_L \eta \ .
\ee
This allows an unambiguous identification of the solution of the nilpotency constraint as $X^1= {(\Omega^1)^2\over \sqrt 2 F^1}$ which was used above. Meanwhile at the supergravity level,  the fermion of the corresponding multiplet transform as 
\be
\delta \chi^1 = {1\over \sqrt 2} P_L (\hat {\slashed {\partial}} z^1 - e^{K/2} g^{1 \bar \beta} \overline \nabla _{\bar \beta} \overline W) \epsilon + {\rm cubic \, in\,  fermions\,  terms} \ ,
\ee
as shown in eq. (18.22)  in \cite{Freedman:2012zz}. We may, therefore,  identify the on-shell value of $F^1$ as $F^1_G$ with $-e^{K/2} g^{1 \bar \beta} \overline \nabla _{\bar \beta} \overline W+$ fermions and proceed with all steps described in the superconformal case. These include the analysis of  i) the dependence in $e^{K/2} g^{1 \bar \beta} \overline \nabla _{\bar \beta} \overline W+$ fermions on $z^1$ and $\bar z^{\bar 1}$ and ii) the fact that the moduli space metric is field-dependent and non-diagonal in direction 1 and the rest of chiral matter. 

In particular, 
we may exclude the terms linear in 
$z^1$ in $F_G^1$ as they originate from the S-supersymmetry preserving gauge-fixing. The same was done in the superconformal model where we have also noticed that the shift of $F^1$ by $ a X^1$ does not affect the solution $X^1= {\Omega^2\over 2F^1} = {\Omega^2\over 2(F^1 + a X^1)}$. Thus, if we would like to associate $e^{K/2} g^{1 \bar \beta} \overline \nabla _{\bar \beta} \overline W$+ fermions  with $-F^1_G$, we may ignore terms proportional to $z^1$ but not proportional to $\bar z^{\bar 1}$. The terms in $e^{K/2} g^{1 \bar \beta} \overline \nabla _{\bar \beta} \overline W$ which depend on $\bar z^1$ are of a different nature. But such terms will be contracted with $ \bar F^{\bar 1}$. Therefore the relevant terms proportional to $\bar z^{\bar 1}$ will be  multiplied by $ \bar F^{\bar 1}$ which will make them  $\bar F$-independent, as we have observed in the superconformal case above. Thus we may identify $ e^{K/2} g^{1 \bar \beta} \overline \nabla _{\bar \beta} \overline W+$ fermions taken at $z^1= \bar z^{\bar 1}=0$ with the relevant part of the auxiliary field of the nilpotent multiplet. In this process one should carefully identify all other terms in the supergravity action where the chiral nilpotent multiplet is taken off shell, so that the total action is not the one in sec. 18.1 in \cite{Freedman:2012zz}, but its partially off shell version with regard to a nilpotent multiplet. In such case we would arrive to the action of the form shown in eq. \rf{actionSC4} with $X$ replaced by $z$. If we would ignore that  $z= {\chi^2\over 2F}$ and integrate out $F$ for $F$-independent $z$ we will get the standard supergravity in sec. 18.1 in \cite{Freedman:2012zz}. But in case we take into account the nilpotency condition, starting with eq. \rf{actionSC4} we will find an action with  non-linear terms in $\chi$ given in eq. \rf{actionSC5}.

In practical terms to find a complete action for matter coupled supergravity means to find the explicit expressions for the entries into action \rf{actionSC4}  for $F_G, A,B,C$ for specific $K(z, \bar z)$, $W(z)$ and $f_{AB }$. Once these are known, the complete non-linear in fermions action is given in 
eq. \rf{actionSC6}. At this point no further steps are required: all local symmetries like the Weyl and $U(1)$ and special superconformal symmetry are already fixed.

Therefore it appears advantageous to use the strategy developed above directly in the supergravity setting.

\section{Local Supersymmetry Gauge-fixing}

In supergravity interacting with matter multiplets, including a nilpotent one, the action has the following terms mixing gravitino $\psi_\mu$ with goldstino $v$,  the combination of other fermions
\be
 \bar \psi^\mu \gamma_\mu \,  v +h.c.= \bar \psi^\mu \gamma_\mu  \left[  {1\over \sqrt 2}  e^{K\over 2} ( \chi^i D_{i }W +  \psi_s D_s W)+ {1\over 2} iP_L \lambda^A {\cal P}_A\right] +h.c.\ .
\ee
Here $\psi^s$ is a VA fermion,  $\chi^i $ are fermions from chiral matter multiplets and $\lambda^A$ are gaugini, using notation in \cite{Freedman:2012zz}. All matter fermions transform under local supersymmetry without a derivatives on $\epsilon(x)$, including the VA fermion \cite{Bergshoeff:2015tra}. Therefore any gauge algebraic in fermions, which does not involve the gravitino, is a unitary gauge since it requires no ghosts. The preferable gauge from the point of view of the absence of mixing of gravitino with other fermions at the minimum of the potential  is the gauge where the goldstino $v$ is vanishing 
\be
v={1\over \sqrt 2}  e^{K\over 2} ( \chi^i D_{i }W +  \psi_s D_s W)+ {1\over 2} iP_L \lambda^A {\cal P}_A=0 \ .
\label{v}\ee
However in models with matter multiplets this gauge, in general, for ${\cal P}_A\neq 0$ and $D_{\alpha }W\neq 0$ at the minimum of the potential, leaves us with an extremely complicated action, with high level of non-linearity on fermions originating from the VA theory. The advantage of the gauge $v=0$ is that gravitino is not mixed with other fermions of the theory.  The disadvantage is that, in case that $D_{i }W\neq 0$ and 
${\cal P}_A\neq 0$ at the vacuum, the action is very complicated since in such case
\be
\psi_s=- {1\over D_s W} \Big (\chi^i D_{i }W +  {1\over 2} iP_L \lambda^A {\cal P}_A\Big) \ ,
\ee
and the action as a function of  independent spinors $\chi^i$ and $\lambda^A$ has many higher order fermion couplings.

Meanwhile, the possible choice of the unitary gauge 
\be
\psi_s=0
\label{VAg}\ee
in general  simplifies the action significantly due to absence of fermion terms beyond quartic, even though there is a gravitino-fermion mixing for models with some $D_{i} W\neq 0$ and ${\cal P}_A\neq 0$ at the minimum. This problem can be taken care by the change of the basis to decouple  $\chi$ and $\lambda$ from  $\bar \psi^\mu \gamma_\mu$.

In this gauge the complete matter-coupled supergravity with chiral and vector multiplets and one nilpotent chiral multiplet, has the following action. First, we note that the original action for all unconstrained multiplets is defined for example in sec. 18.1 in \cite{Freedman:2012zz}, let us call it $e^{-1} {\cal L}^{\rm book}$. It depends on $e_\mu^a,\psi_\mu,z^\alpha,\bar{z}^{\bar \alpha}, \chi^\alpha,\chi^{\bar \alpha},A_\mu^A,\lambda^A$. The complete action with a local supersymmetry is complicated due to a non-Gaussian nature of the auxiliary field $F^1$ of the nilpotent multiplet $(z^1=s, \chi^1=\psi_s, F^1= F_s)$. However, in the unitary gauge \rf{VAg} this feature of the theory disappears: all terms with a non-Gaussian dependence on $F^1$ enter via ${\psi_s^2\over 2 F_s}$ and vanish in this gauge. The complete action, defined by $K$ given in \rf{K} and $W, f_{AB} $ given in \rf{W} has to be constructed according to the standard rules. It is
 given for example in the book \cite{Freedman:2012zz} in equations (18.6)-(18.19) and takes the form 
\bea\label{eq:Lb}
&&e^{-1}{\cal L}^{\rm book} \Big ( K(z^i, \bar z^{\bar i}; S, \bar S),  W(z, S) \,  f_{AB}(z, S)\Big ) \Rightarrow \cr 
\cr
&&e^{-1}{\cal L}^{\rm book}(e_\mu^a,\psi_\mu,z^i,\bar{z}^{\bar i}, s, \bar s, \chi^i,\chi^{\bar i}, \psi_s, \bar \psi_s,A_\mu^A,\lambda^A) \ .
\eea
The locally supersymmetric  action with the $S$ multiplet nilpotent has $\psi_s$-dependent terms which are not present in the standard action. However, all these terms are absent in the unitary gauge \rf{VAg} where the physical fields of the $S$ multiplet are absent. Therefore the complete matter-coupled supergravity action with a nilpotent multiplet in the gauge \rf{VAg} is given by 
\be\label{eq:U}
e^{-1} {\cal L}_{\rm unitary} = e^{-1} {\cal L}^{\rm book}(e_\mu^a,\psi_\mu,z^i,\bar{z}^{\bar i}, s, \bar s, \chi^i,\chi^{\bar i}, \psi_s, \bar \psi_s, A_\mu^A,\lambda^A)|_{s=\psi_s=0} \ .
\ee
Despite the physical fields of the nilpotent multiplet, $s$ and $\psi_s$, are both absent in the unitary gauge, the auxiliary field $F_s=-e^{K/2} g^{1 \bar \beta} \overline \nabla _{\bar \beta} \overline W+$ fermions  is still present as shown in eq. \rf{pot} and does affect the action in the unitary gauge. The complete action is different from the one in which the nilpotent multiplet is absent.

For example, there are extra positive terms in the potential due to $|F_s|^2 = |D_SW|^2$. There are additional terms in the mass formula of physical fermions due to terms with $D_SW$ and in case that $f_{AB}$ depends on $S$. However, all of these terms are included in the standard action \rf{eq:Lb}, which is computed in a way that a summation over all chiral multiplets, including $z^1=s$, is performed in the derivation of the action.

The unitary gauge \rf{VAg} is useful and convenient in case that at the vacuum $D_{i} W=0$ and only $D_sW\neq 0$ and the Killing potentials are absent,  ${\cal P}_A=0$.  In the case of a single inflaton chiral multiplet $\Phi$, present in the theory in addition to a nilpotent one, the condition $D_\Phi W=0$ is fulfilled in inflationary models constructed in most models in \cite{Kallosh:2014via,Dall'Agata:2014oka},  where $\Phi$ is the inflaton multiplet. When more general matter multiplets $z^i$ are present, it is possible to provide this condition taking canonical \K\, potential $z^i\bar z^{\bar i}$ and quadratic superpotential $(z^i)^2$ and providing the minimum at $z^i=0$, as suggested in \cite{Dall'Agata:2014oka}. Thus for such models the gauge \rf{VAg} is very useful.  

In general, some other classes of gauges fixing local supersymmetry may be also useful. We postpone a discussion of other gauges for future studies. Such studies  can be performed after the explicit form of the complete locally supersymmetric action will be derived in \cite{RT}, following the strategy proposed in this paper.

\section{Conclusion}

We have found here that it is possible to include the nilpotent multiplet in the general supergravity models and to derive the complete locally supersymmetric action, including fermions.   Our main result is shown in eq. \rf{actionSC5} and it is explained in the paper why this action is valid for matter coupled supergravity. The elimination of auxiliary fields  
from the action despite their non-Gaussian dependence can be done using the same procedure as in the case of pure de Sitter supergravity \cite{Bergshoeff:2015tra}.  An explicit realization of  this procedure and construction of  such general de Sitter type supergravities will be presented in \cite{RT}. It will require us to find the actual values of the symbolic expressions in \rf{actionSC5} which would be valid for  general matter-coupled supergravities. But when they are known, the result follows either in the form \rf{actionSC5}, or in the form \rf{actionSC6}. A particular feature of this answer is that all higher order fermion terms in this locally supersymmetric action with the non-linearly realized supersymmetry are given in the closed form.

We have also observed that in the unitary gauge, where the fermion from the nilpotent multiplet vanishes, the  full  action is very simple and given in eq. 
\rf{eq:U}. Fermion  couplings are not higher than quartic in this gauge. However,  other interesting gauges are possible with  some complementary properties.
It would be interesting to derive the action in other gauges, when the detailed locally supersymmetric action will be available, and to study the physics of these models with non-linearly realized local supersymmetry. 

The class of models with general matter-coupled supergravity and de Sitter vacua associated with a  non-linearly realized supersymmetry  lead  to interesting phenomenology both in cosmological setting as well as in particle physics \cite{Kallosh:2014via,Dall'Agata:2014oka}. These models have  a potential for explaining dark energy, inflation and susy breaking. It is therefore interesting to find out that such theories have a rather universal form for a general matter coupling.

\section*{Acknowledgments}

We are grateful  to   E. Bergshoeff, J.~J.~M.~Carrasco,  K. Dasgupta,  S. Ferrara, D. Freedman, A. Linde,  F. Quevedo, A. Uranga,  and especially to  A. Van Proeyen and T. Wrase, for stimulating  discussions and for the collaboration of related projects. 
This work 
is supported by the SITP,   by the NSF Grant PHY-1316699 and  by the Templeton foundation grant `Quantum Gravity Frontiers'.


\end{document}